# Evidence for studying interactions between science and policy: An exploration of scholarly and policy references in Overton-indexed policy documents

Biegzat Murat, Zhichao Fang*, Ed Noyons, Rodrigo Costas

* Corresponding author: z.fang@cwts.leidenuniv.nl

Biegzat Murat (ORCID: 0000-0001-7828-2472)

1. Centre for Science and Technology Studies (CWTS), Leiden University, Leiden, The Netherlands.

Zhichao Fang (ORCID: 0000-0002-3802-2227)

1. School of Information Resource Management, Renmin University of China, Beijing, China.
2. Centre for Science and Technology Studies (CWTS), Leiden University, Leiden, The Netherlands.
E-mail: z.fang@cwts.leidenuniv.nl

Ed Noyons (ORCID: 0000-0002-3103-2991)

1. Centre for Science and Technology Studies (CWTS), Leiden University, Leiden, The Netherlands.

Rodrigo Costas (ORCID: 0000-0002-7465-6462)

1. Centre for Science and Technology Studies (CWTS), Leiden University, Leiden, The Netherlands.
2. DSI-NRF Centre of Excellence in Scientometrics and Science, Technology and Innovation Policy, Stellenbosch University, Stellenbosch, South Africa.

**Abstract:**

[Purpose] Overton, a global policy index, provides new opportunities to study the interactions between science and policy. This study aims to characterize the presence of scholarly and policy references in Overton-indexed policy documents and examine their distribution across key bibliographic dimensions, thereby assessing Overton's potential as a data source for policy metrics.

[Design/methodology/approach] We analyze a dataset of approximately 17.5 million policy documents from Overton, incorporating metadata such as publication year, policy source, country, language, subject area, and policy topic. Descriptive statistics are employed to assess the presence and distribution of reference data across these dimensions.

[Findings] Overton indexes a substantial volume of policy documents and identifies considerable reference data within them: 7.7% of documents contain scholarly references and 10.6% contain policy references. However, the presence of references varies significantly across publications years, source types, countries, languages, subject areas, and policy topics, indicating coverage biases that may affect interpretations of policy impact.

[Research limitations] The analysis is based on the Overton database as of June 2025. As Overton is regularly updated, the distribution patterns of indexed documents and references may evolve over time.

[Practical implications] The findings offer insights into the opportunities and constraints of using Overton for investigating evidence-based policymaking and for assessing the policy uptake of research outputs in the context of research evaluation.

[Originality/Value] This is the first large-scale study to systematically examine the distribution of reference data in Overton. It contributes a foundational understanding of this emerging source for policy metrics, highlighting both its potential applications and limitations, and underlining the importance of addressing current coverage imbalances.

## 1. Introduction

The call for evidence-based policymaking is proliferating across all domains of public service (Black, 2001). One of its core principles involves the utilization of "the best available research findings at all stages of the policymaking process" (Kolahi & Khazaei, 2018). This underscores the critical role of research evidence in supporting the development and implementation of policy initiatives (Sanderson, 2002). Conversely, within the scientific context, informing policy and influencing decision-making are regarded as among the most significant aspects associated with the broader *societal impact of science* (Gauch & Blümel, 2016). Wilsdon *et al.* (2015) define the societal impact of scientific research as "auditable or recorded influence achieved upon non-academic organization(s) or actor(s) in a sector outside the university sector itself", which needs "to be demonstrated rather than assumed". In the context of policymaking – and more specifically, "science-based policymaking" (Pedersen, 2014) – scholarly references embedded within policy documents can offer direct evidence of the relevance of scientific research to policymaking. Therefore, policy documents and their citations to scientific results are considered among the most pertinent sources for measuring the societal impact of research outputs (Bornmann et al., 2016).

*1.1. Altmetric studies related to policy documents*

Altmetrics broaden the scope of the impact made by scientific research, enabling more diverse forms of impact analysis (Waltman & Costas, 2014) and facilitating advanced quantitative studies of science-society interactions (Costas et al., 2021). Since Altmetric – one of the leading altmetric data aggregators – incorporated policy documents as one of its data sources, numerous quantitative studies have investigated policy citations of research outputs and attempted to measure the impact of scientific knowledge on policy.

Most previous studies have explored the presence of policy citations across research outputs in various contexts. Large-scale and cross-disciplinary analyses have confirmed that only a very limited share of scientific publications has been cited by the policy documents recorded by Altmetric. For instance, Haunschild and Bornmann (2017) found that in a set of nearly 11.3 million Web of Science (WoS) papers published between 2000 and 2014, only 0.32% had at least one policy citation detected by Altmetric. A slightly higher coverage was reported in an extensive analysis by Fang *et al.* (2020), which concluded that for nearly 12.3 million WoS-indexed papers published between 2012 and 2018, about 1.12% had been cited by policy documents at least once.

Despite the overall low presence of policy citations among scientific outputs, these citations can still be useful in identifying research outputs with above-average impact in policy documents (Noyons, 2019; Thelwall et al., 2013). The extent to which a research output is cited by policy documents can reflect its relevance in the policy sphere. Thus, policy citations detected by Altmetric are seen as a valuable form of evidence to measure the relevance of research outputs in societal areas specifically relevant to policy (Bornmann et al., 2016). Other aspects studied regarding policy citations to research outputs include the evaluation of the policy impact of publications by a specific university (Tattersall & Carroll, 2018), the verification of the societal impact of publications measured by altmetric indicators (e.g., policy citations, news media mentions, and social media mentions) in contrast to the assessment results by some national evaluation systems and funders (Bornmann et al., 2019; Kassab et al., 2020), the comparison of policy citations received by scientific publications with different open access statuses (Taylor, 2020), and the measurement of the policy impact of research in specific policy-relevant fields such as climate change and economics (Bornmann et al., 2016; Drongstrup et al., 2020). Moreover, some studies have paid close attention to critical issues that might affect the measurement of policy impact, such as "citation delays" (Fang & Costas, 2020), citing motivations (Yu et al., 2023), and the data quality of policy citations in Altmetric (Tattersall & Carroll, 2018; Yu et al., 2020).

*1.2. Overton as an emerging data source for policy metrics*

In 2019, the launch of Overton (https://www.overton.io/), a searchable index of policy documents, opened new possibilities for monitoring science-based policymaking and tracking the policy impact of research outputs. According to Overton (2024a), it provides the largest existing global index of policy documents. Overton broadly defines policy documents as "documents written primarily for or by policymakers that are published by a policy-focused source" (Overton, 2024i), which includes not only policymakers from governments (local, regional, national, or supranational) but also other types of organizations such as think tanks, non-governmental organizations (NGOs), and intergovernmental organizations (IGOs). Szomszor and Adie (2022) have presented a comprehensive analysis of the composition of policy documents indexed in Overton.

As shown in Figure 1, Overton provides a wide range of bibliographic information on policy documents, including title, publication date, source, and topics, as well as citation relationships detected through text-mining solutions (Overton, 2024d). The citation relationships involving Overton-indexed policy documents consist of both the citations they receive from other indexed policy documents and the references they make to research. This makes it possible to uncover the citation interactions between policy and science (i.e., policy-to-science citations) and among policy documents themselves (i.e., policy-to-policy citations).

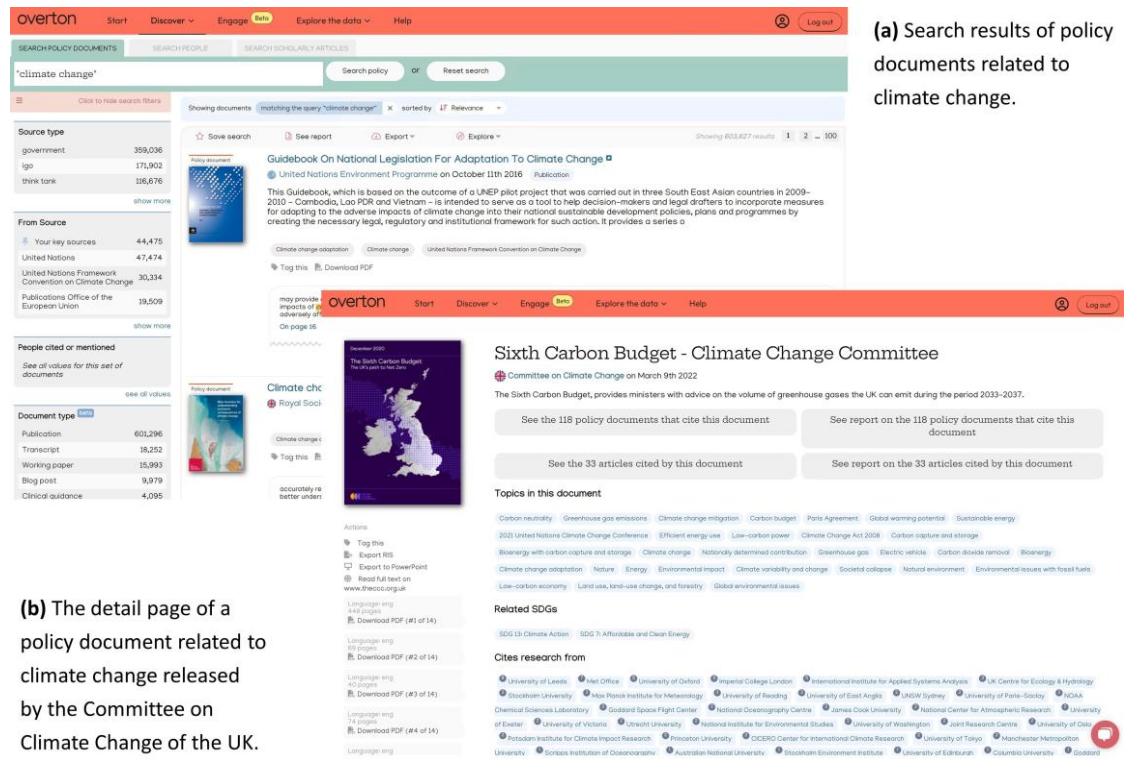

**Figure 1**. Screenshots of the Overton web interface: (a) an example of search results and (b) a detail page of a specific policy document.

Since its release, Overton has been widely adopted by researchers to evaluate the policy impact of research on a variety of topics, such as public policy (Dorta-González et al., 2024), scientometrics (Atapour et al., 2024), interdisciplinarity (Pinheiro et al., 2021), and the COVID-19 pandemic (Adie, 2020; Hu et al., 2024; Yin et al., 2021). Additionally, Overton is used to assess the policy visibility of individual academics (Jonker & Vanlee, 2024). In these studies, citations to scholarly papers in policy documents provided by Overton are leveraged to quantify the extent to which research outputs have attracted attention from policymakers (or at least from authors of policy documents). These studies have greatly contributed to understanding the potential of Overton in revealing the interactions between science and policy. However, as an emerging data source for policy metrics, it remains important to examine the distribution of reference data in Overton-indexed policy documents and its potential limitations caused by the distribution of reference data.

*1.3. Objectives of the study*

Given Overton's potential to advance the study of evidence-based policymaking and to support empirical research on how science impacts policymaking, it is critical to develop a comprehensive understanding of the main characteristics of the different metadata elements indexed in Overton. More importantly, unraveling the potential influence of the distribution of reference data on interpreting results based on Overton is essential.

The main objective of this study is to characterize the presence of scholarly references and policy references in Overton-indexed policy documents. To achieve this objective, we address the following research questions (RQs):

- **RQ1**. How are scholarly and policy references distributed across Overton-indexed policy documents with respect to publication year?

- **RQ2**. How are scholarly and policy references distributed across Overton-indexed policy documents with respect to policy source?

- **RQ3**. How are scholarly and policy references distributed across Overton-indexed policy documents with respect to geographic origin?

- **RQ4**. How are scholarly and policy references distributed across Overton-indexed policy documents with respect to language?

- **RQ5**. How are scholarly and policy references distributed across Overton-indexed policy documents with respect to subject area?

- **RQ6**. How are scholarly and policy references distributed across Overton-indexed policy documents with respect to policy document topics?

**2. Data and methods**

*2.1. Dataset of policy documents*

A total of 17,522,124 policy documents were sourced from the snapshot of the Overton database (version dated June 2025), which is maintained by the Centre for Science and Technology Studies (CWTS) at Leiden University. A diverse range of Overton-indexed metadata elements was analyzed to characterize the distribution of policy documents and to assess the presence of both scholarly and policy references. Tables 1 and 2 outline the bibliographic and reference metadata elements provided by Overton that were included in this study, respectively.

**Table 1**. Bibliographic metadata elements from Overton-indexed policy documents

| Metadata type | Definition |
| --- | --- |
| Publication date | The date when the policy documents were published. |
| Policy source | The organizations or entities from which policy documents are collected. |
| Source type | The categories of policy sources. Overton classifies policy sources into three main types: "government", "think tank" and "intergovernmental organization (IGO)" (Overton, 2024h). Additionally, Overton tracks policy documents from "other" sources, including open repositories and non-governmental organizations (NGOs). |
| Source country | The countries associated with the policy sources. Notably, IGOs and the European Union (EU) are listed separately. |
| Document language | The languages in which the PDF files of policy documents are written. A single policy document may be released in multiple PDF versions across different languages.[1] |
| Subject area | The subject areas of policy documents. For documents written in certain languages (e.g., English, French, Spanish), Overton assigns subject areas by matching phrases and entities extracted from the full text with examples from each category in the International Press Telecommunications Council (IPTC)'s Media Topics controlled vocabulary (https://iptc.org/standards/media-topics/) [2] (Overton, 2024a). |
| Topics | The main themes of policy documents. Overton assigns topics to documents written in certain languages (e.g., English, French, Spanish) by matching phrases and entities extracted from the full text with the titles of Wikipedia pages (Overton, 2024a). |

---

[1] In our dataset, a total of 17.5 million distinct policy documents have been released in over 19.6 million PDF files.

[2] This taxonomy is primarily used by newspapers and magazines to organize their articles. The full taxonomy of Media Topics can be accessed at:
https://www.iptc.org/std/NewsCodes/treeview/mediatopic/mediatopic-en-GB.html.

**Table 2**. Reference metadata elements from Overton-indexed policy documents

| Metadata type | Definition |
|---|---|
| Scholarly references | Citations from policy documents to academic literature (i.e., scholarly papers referenced within policy documents). Overton identifies and formats these references by extracting elements from potential reference strings - such as sources, titles, and publication years - from the full text of policy documents, then searching Crossref to retrieve the DOIs of the cited scholarly works (Overton, 2024d, 2024c). |
| Policy references | Citations from one policy documents to another (i.e., policy documents referenced within other policy documents). Overton identifies and formats these references using a method similar to that employed for scholarly references. Policy references are obtained by matching elements of potential reference strings extracted from the full text with indexed policy documents in the Overton database (Overton, 2024d, 2024c). |

In this study, we focus specifically on the presence of both scholarly and policy references within policy documents detected by Overton. To ensure that the DOIs identified by Overton as references in the indexed policy documents correspond to valid research outputs, we cross-referenced all cited DOIs with four in-house academic databases at CWTS: the Web of Science (version dated March 2025), Scopus (version dated March 2025), Dimensions (version dated July 2024), and OpenAlex (version dated August 2024). Only the cited DOIs indexed in at least one of these four academic databases were considered as scholarly references in this study. Detailed coverage of the cited DOIs in these databases is provided in Table A1 in Appendix A.

*2.2. Indicators and analytic approaches*

Table 3 presents the descriptive statistics of scholarly and policy references within policy documents. Both types of references exhibit highly skewed and zero-inflated distributions, as indicated by the percentile values as well as the extreme skewness and kurtosis statistics. Specifically, approximately 92.3% of policy documents contain no scholarly references, while 89.4% contain no policy references, highlighting the zero-inflation in the reference data of policy documents.

**Table 3**. Descriptive statistics of scholarly and policy references in policy documents

| Indicator | Scholarly references | Policy references |
|---|---|---|
| Minimum | 0 | 0 |
| Maximum | 14,633 | 1,352 |
| 25th percentile | 0 | 0 |
| 50th percentile (median) | 0 | 0 |
| 75th percentile | 0 | 0 |
| 90th percentile | 0 | 1 |
| 99th percentile | 32 | 9 |
| Arithmetic mean | 1.36 | 0.45 |
| Standard deviation | 17.47 | 3.22 |
| Skewness | 154.16 | 46.51 |
| Kurtosis | 67,556.53 | 8,123.57 |

Given the highly skewed distribution of both scholarly and policy references, we employ three indicators to measure their presence within policy documents: *total*, *coverage*, and *average*, as detailed below:

- Total refers to the sum of scholarly or policy references within a particular set of policy documents. This metric reflects the absolute volume of references, highlighting the *primary contributors* of scholarly or policy references in Overton.

- Coverage denotes the percentage of policy documents containing at least one scholarly or policy reference. This indicator represents the *breadth* of the presence of scholarly or policy references across the dataset.

- Average represents the mean number of scholarly or policy references within a given set of policy documents. Due to the highly skewed and zero-inflated nature of the reference distribution, we utilize the geometric mean instead of the arithmetic mean, as recommended by Fairclough and Thelwall (2015) and Thelwall (2016). The log-transformed geometric mean (Thelwall, 2015; Thelwall & Fairclough, 2015) is calculated as follows:

$$Average = \exp\left(\frac{\sum_{i=1}^{n} \ln(R_i + 1)}{n}\right) - 1$$

where $n$ denotes the number of policy documents in the dataset, and $\ln(R_i + 1)$ represents the natural logarithm of the number of scholarly or policy references in the $i_{th}$ policy document. Adding one to the number of references addresses the issue of policy documents with zero references. After applying the exponential function, one is subtracted to adjust the result. The average thus reflects the *intensity* of scholarly or policy references within a given set of policy documents.

For the overall dataset consisting of approximately 17.5 million policy documents, Table 4 presents the results of the three indicators measuring the presence of scholarly and policy references. Overall, Overton identifies a larger number of scholarly references than policy references within the indexed policy documents. Policy references exhibit wider coverage, as reflected by their relatively higher coverage value, while the distribution of scholarly references is more intensive, as evidenced by the higher average.

**Table 4**. Total, coverage, and average of references for the overall dataset

| Indicator | Scholarly references | Policy references |
|---|---|---|
| Total | 23,745,211 | 7,806,381 |
| Coverage | 7.7% | 10.6% |
| Average | 0.17 | 0.14 |

## 3. Results

This section comprises five subsections that explore the presence of both scholarly and policy references within Overton-indexed policy documents. These aspects are analyzed across publication years, policy sources, countries, languages, subject areas, and policy topics. Each dimension provides distinct insights into the relevance and limitations of Overton-indexed policy documents for various analytical purposes.

*3.1. Presence of references over publication years*

Figure 2 illustrates the distribution of scholarly and policy references over time based on the publication years of the policy documents. Overall, Overton primarily indexes recent policy documents published from 2009 onwards, resulting in lower coverage of older documents (Overton, 2024e). Consequently, more recent policy documents account for most scholarly and policy references due to their volume advantage. However, when considering both coverage and average, scholarly references tend to be more prominent in older policy documents compared to more recent ones. This suggests a relatively weaker breadth and intensity of scholarly references in newer policy documents. In contrast, policy references are more prevalent in recent policy documents, possibly because Overton identifies them by matching with previously indexed policy documents. The increase in the number of recent policy documents enhances the likelihood of detecting policy references.

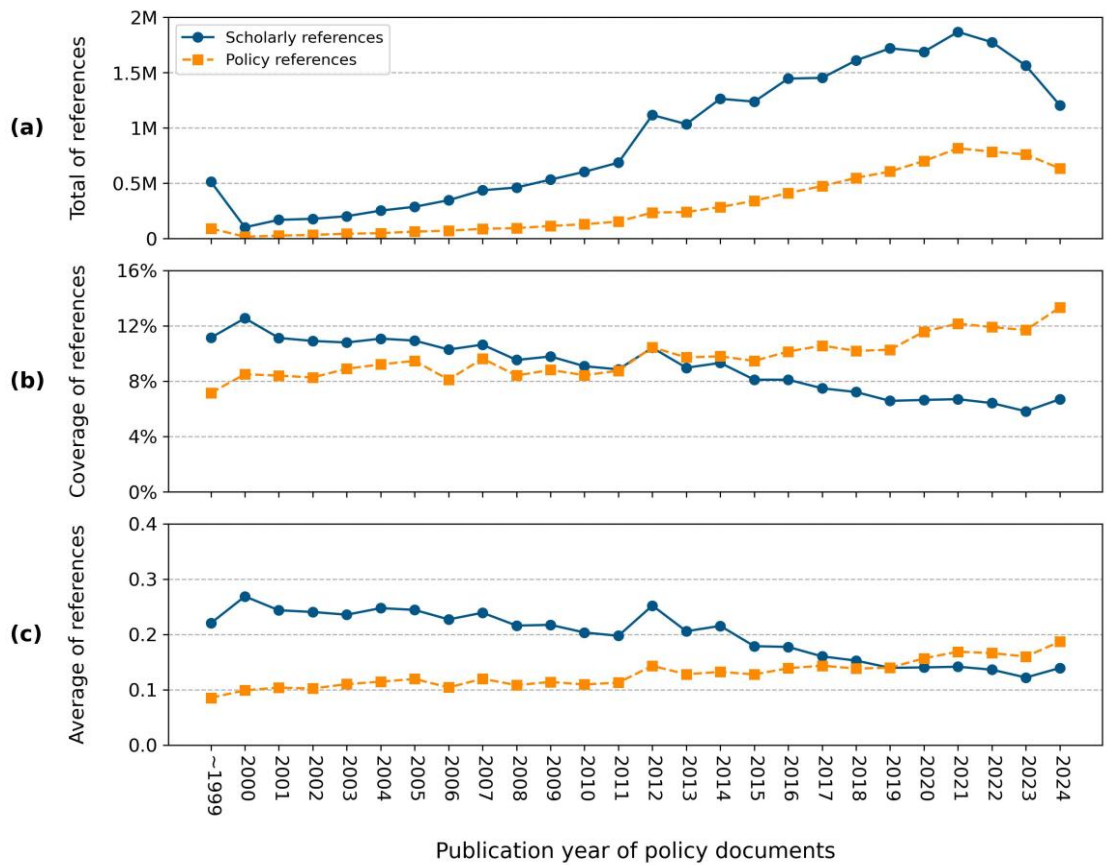

**Figure 2**. Presence of references in policy documents over publication years, presented through three perspectives: (a) total, (b) coverage, and (c) average.

*3.2. Presence of references across policy sources*

Within our dataset, Overton tracks policy documents from 2,534 distinct sources, categorized into four main types: "government", "think tank", "IGO", and "other" sources such as open repositories or NGOs. Figure 3 illustrates the composition of Overton-indexed policy documents based on these source types, along with the top 10 individual sources within each category contributing the highest number of documents. Among these sources, 1,136 are governments, whose policy documents account for 85.7% of the total indexed by Overton. This is followed by IGO sources, totaling 125, which contribute 8.0% to the overall database. Think tanks, represented by 1,244 sources, contribute 6.0% of the policy documents. The remaining 0.3% originate from other sources, primarily repositories such as the *Analysis & Policy Observatory* and the

*Guidelines in PubMed Central.*[3]

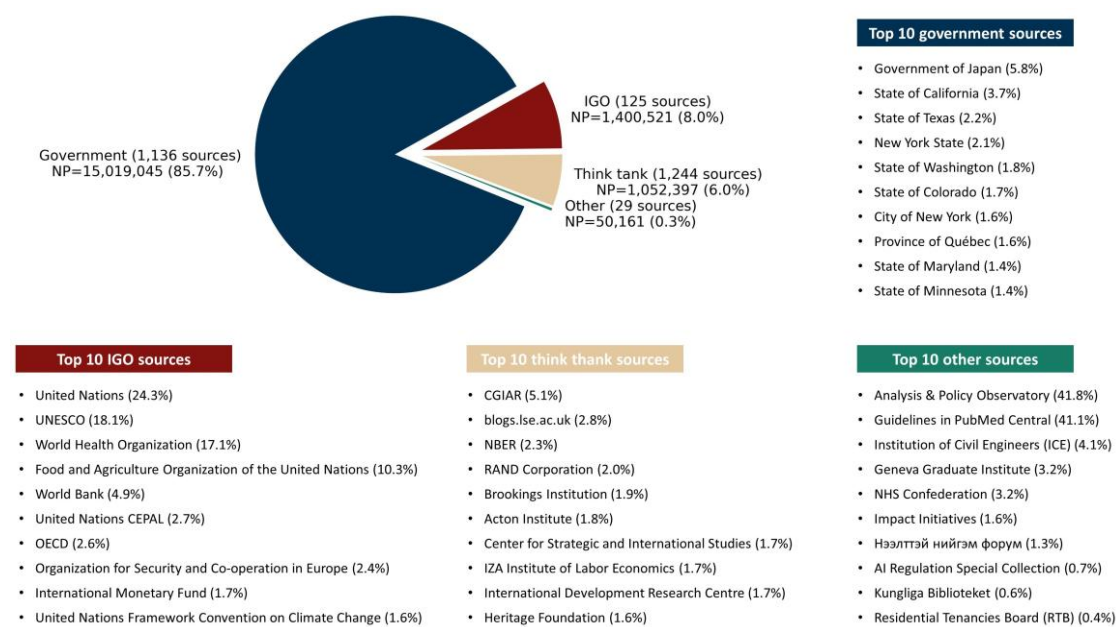

**Figure 3**. Distribution of policy documents across the four source types, and the top 10 policy sources contributing the highest number of documents within each source type. For each of the top 10 sources, the proportion of policy documents relative to the total for that source type is also presented.

Figure 4 illustrates the presence of references in policy documents across four policy source types. Government sources, which account for the largest share of policy documents indexed by Overton, also contribute the greatest total number of scholarly references. However, only 4.7% of government-published policy documents include at least one scholarly reference. By contrast, although the other three source types have fewer total scholarly references, they exhibit higher coverage and average numbers of scholarly references in their documents. Specifically, 20.9% of IGO-published policy documents and 31.0% of think tank documents cite at least one scholarly reference. Policy documents from other open repositories or NGOs have the highest coverage, with 68.3% including scholarly references; however, these account for only about 6.1% of the total scholarly references recorded in Overton. With respect to intensity, government documents have the lowest average number of scholarly references, largely due to the predominance of documents without citations to research. In contrast, think tank and IGO documents show higher averages, while those from other repositories or NGOs exhibit the highest intensity of scholarly references. As Overton increasingly indexes government policy documents over time (see Figure B1 in Appendix B), this may partly explain the overall decline in the coverage and average number of scholarly references observed in recent years (Figure 2), since government sources demonstrate a weaker tendency to cite scholarly work in policymaking.

---

[3] The Analysis & Policy Observatory (APO, https://apo.org.au/) is an open-access digital repository for grey literature related to public policy issues, with a primary focus on Australia and New Zealand. Guidelines in PubMed Central collects clinical guidelines published in biomedical and life sciences journals that are available in PubMed Central (PMC, https://www.ncbi.nlm.nih.gov/pmc/), a free full-text archive developed and maintained by the U.S. National Institutes of Health.

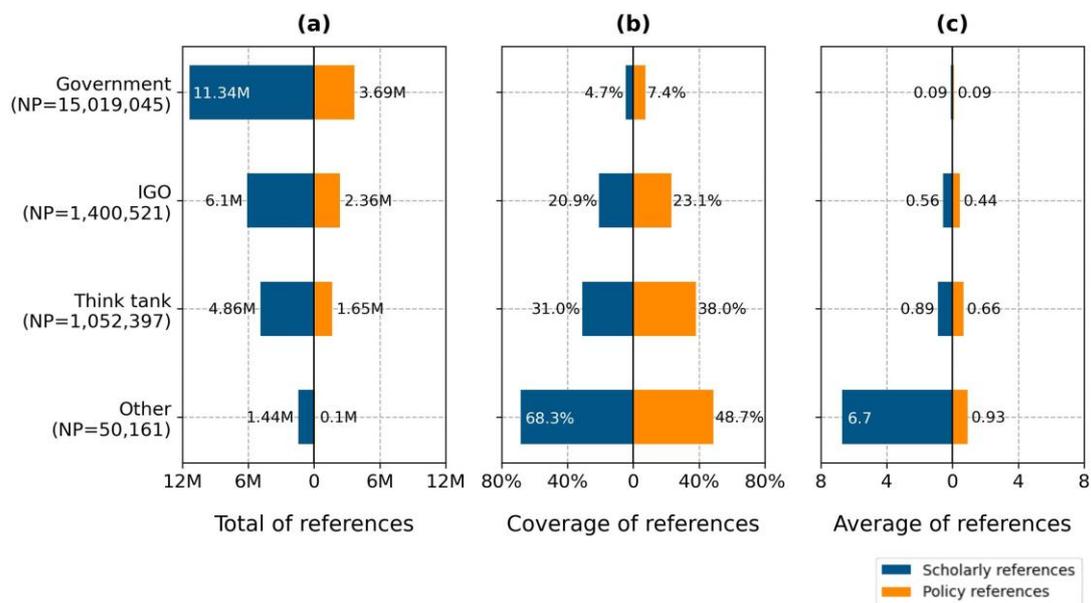

**Figure 4**. Presence of references in policy documents across policy sources, presented through three perspectives: (a) total, (b) coverage, and (c) average.

In comparison to scholarly references, policy references are less common in Overton-indexed policy documents in terms of the total metric. Government sources exhibit the lowest coverage, with only 7.4% of their policy documents citing at least one other policy document. The proportion is higher for the other three source types. For instance, 38.0% of think tank documents contain at least one policy reference, while nearly 48.7% of policy documents from other open repositories or NGOs include policy references. In terms of the average number of policy references per document, the performance across source types is relatively similar, with each citing between 0.1 and 0.9 policy references on average.

*3.3. Presence of references across countries and languages*

According to Overton (2021), the platform indexes "more documents from more developed countries than elsewhere". A large-scale bibliometric study by Szomszor and Adie (2022) mapped the global distribution of Overton-indexed policy documents and confirmed that the database provides substantial coverage for the US, Canada, Japan, and several Western European countries. This tendency is also evident in our dataset, which shows a predominance of policy documents from North America, Europe, and IGOs (see Figure C1 in Appendix C). Consequently, policy documents from different regions contribute to scholarly and policy references to varying extents, as demonstrated in Figure 5. In our dataset, Overton tracks policy sources from 193 countries or regions. For clarity, Figure 5 focuses on the top 10 countries with the highest number of indexed policy documents and depicts their inclusion of scholarly and policy references.

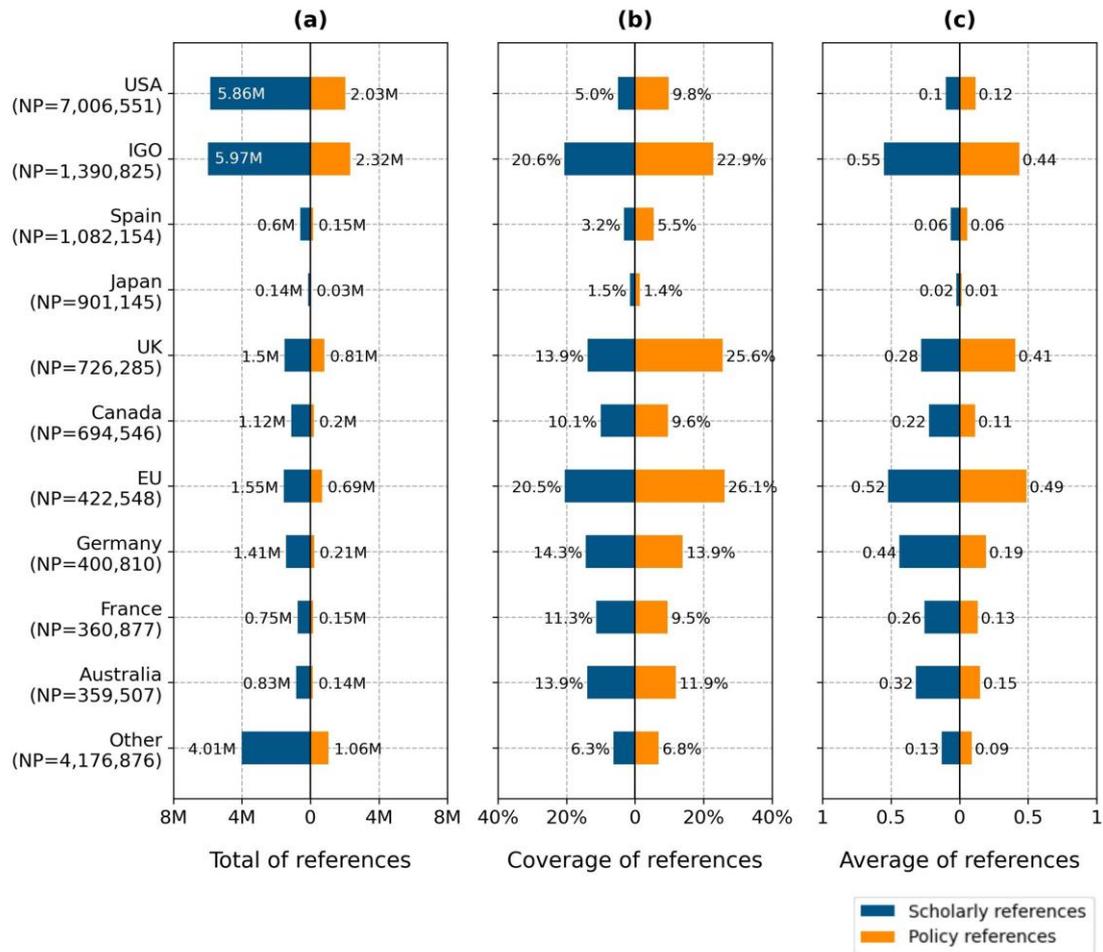

**Figure 5**. Presence of references in policy documents from the top 10 countries/regions with the highest number of policy documents, presented through three perspectives: (a) total, (b) coverage, and (c) average.

From a total volume perspective, the US and IGO contribute the most scholarly and policy references, largely due to their higher number of indexed policy documents. Other countries that significantly contribute to scholarly and policy references are primarily from Europe and North America. In contrast, countries in Asia, Africa, and Latin America are underrepresented in Overton, with fewer scholarly and policy references originating from these regions. An exception is Japan, which ranks fourth after the US, IGO, and Spain in terms of the number of Overton-indexed policy documents. However, policy documents from Japan contain relatively fewer scholarly and policy references compared to those from other leading countries.

In terms of coverage and average, IGO and the EU show broader representation, with their policy documents containing more intensive references among the top countries.

The US, despite accounting for 40.0% of Overton-indexed policy documents and contributing the highest total number of references, has a relatively moderate proportion of documents containing scholarly references (5.0%) and policy references (9.8%).

Based on the language information reported by Overton in the PDF files of policy documents, Overton-indexed policy documents are available in 74 different languages.[4] Figure 6 illustrates the presence of scholarly and policy references in policy documents written in the top 10 most used languages. English dominates, accounting for 67.1% of all indexed policy documents. Other languages represent smaller shares, such as Spanish (9.3%), French (6.0%), and Japanese (4.1%).

Due to the large volume of English-language documents, policy documents in English contribute the overwhelming majority of both scholarly and policy references, making them the predominant source of reference data in Overton. Although documents in other languages are less represented, many exhibits reference levels comparable to those in English in terms of coverage and average. However, policy documents in Japanese display a weaker presence of both scholarly and policy references, consistent with the trends observed in Figure 5.

---

[4] For policy documents that have multiple PDF files in different languages, we applied full counting when calculating the number of policy documents for each language. For example, if a policy document is published in both English and French versions, each language receives a count of one for that policy document.

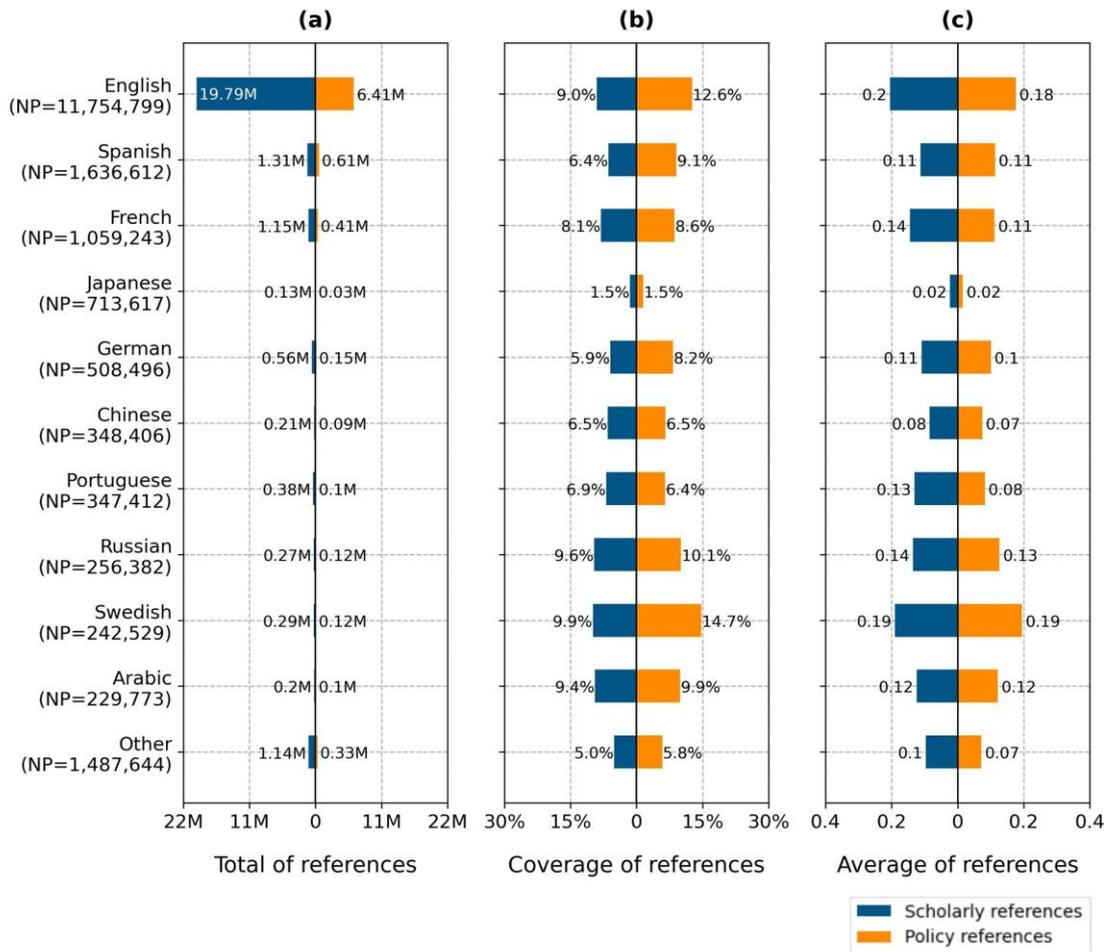

**Figure 6**. Presence of references in policy documents across the top 10 most used languages, presented through three perspectives: (a) total, (b) coverage, and (c) average.

*3.4. Presence of references across subject areas*

Utilizing the subject area information assigned by Overton, Figure 7 illustrates the presence of reference data across 18 subject areas.[5] *Science and technology* is the most common subject area in the Overton database, with 44.4% of indexed policy documents related to this field. The next most represented areas are primarily associated with the economy, politics, society, healthcare, and environment. Generally, subject areas with a larger number of indexed policy documents also exhibit a higher number of references.

---

[5] As with language classifications, a single policy document can be assigned to multiple subject areas. In such cases, full counting was applied when calculating the number of policy documents for each subject area.

However, *Politics* and *Crime, law, and justice* are notable exceptions, displaying relatively few scholarly and policy references despite having a substantial number of policy documents.

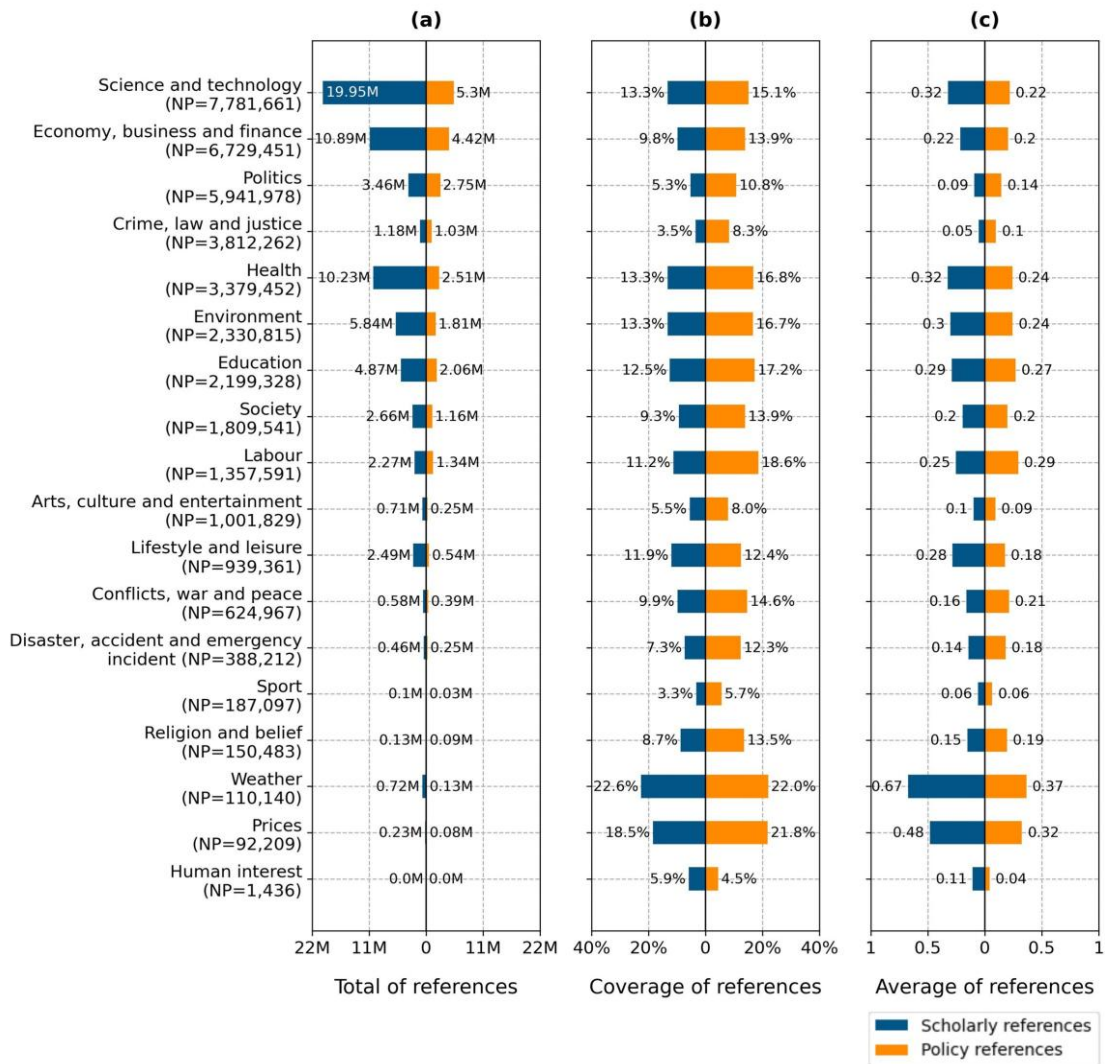

**Figure 7**. Presence of references in policy documents across subject areas, presented through three perspectives: (a) total, (b) coverage, and (c) average.

From the perspectives of coverage and average, policy documents in subject areas associated with science and technology, health, environment, and economy tend to show a higher presence of scholarly references. In contrast, the presence of scholarly references is relatively low in areas related to politics, law, and recreation and sports (e.g., *Politics*, *Crime, law and justice*, *Arts, culture and entertainment*, and *Sport*). A similar pattern is observed for policy references, though with some exceptions. Subject areas associated with political, legal, and religious themes - such as *Politics*, *Crime, law and justice*, *Conflicts, war and peace*, and *Religion and belief* - are more likely to reference other policy documents than scholarly research outputs.

*3.5. Presence of references across policy topics*

We further examined the distribution of Overton-indexed policy documents at the topic level. Figure 8(a) presents a co-word network of policy topics identified by Overton, created using the VOSviewer software. To enhance visualization, only topics occurring in at least 10,000 policy documents were included in the network, resulting in a set of 3,402 topics representing the landscape of Overton-indexed policy documents. These topics form six clusters, which can be broadly labeled as "law", "politics", "economy", "natural environment", "healthcare", and "research & education". The top 10 most frequently occurring topics within each cluster are listed separately to provide better insights into these categories. The six topic clusters are linked to the 18 subject areas to varying extents, with patterns that generally align with the most thematically relevant subject areas of policy documents (see Figure D1 in Appendix D).

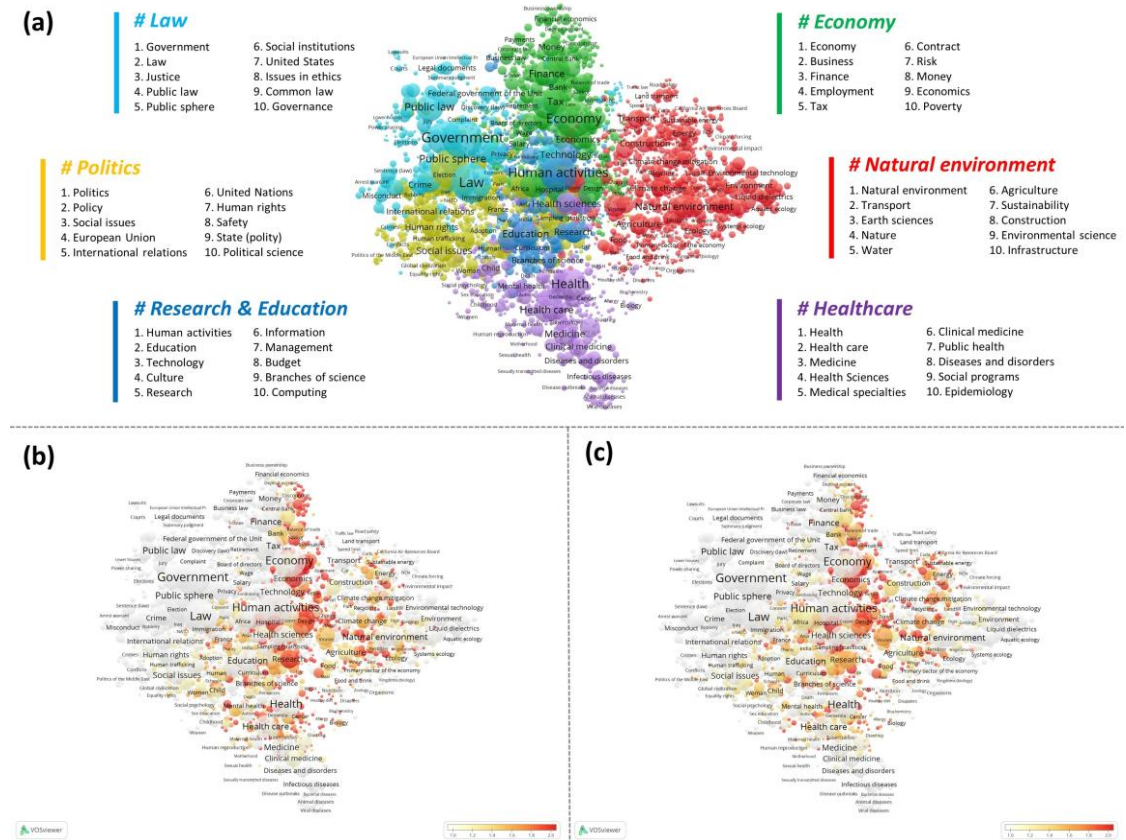

**Figure 8**. (a) Co-word network of topics in Overton-indexed policy documents;

overlay visualizations of topics colored based on scores reflecting (b) the coverage of scholarly references and (c) the coverage of policy references. In the overlay visualizations, scores of topics are normalized using the "divide by mean" function in VOSviewer. The redder a topic node, the higher the coverage of scholarly or policy references in the policy documents related to that topic, relative to all topics in the map.

Using VOSviewer's overlay visualization, Figures 8(b) and 8(c) display the co-word network of policy document topics, scored based on the coverage of scholarly references and policy references, respectively. Figure 8(b) reveals that topics situated in the central and right parts of the network - such as those related to healthcare, natural environment, economy, and some scientific research and education activities - show higher levels of scholarly references in their associated policy documents. In contrast, topics related to politics and law (located in the left part of the network) exhibit lower coverage of scholarly references, indicating that Overton-indexed policy documents on these topics are less likely to cite research outputs. These findings are further validated by the boxplot in Figure 9, which demonstrates the distribution of reference coverage across topics within the six clusters. Policy topics concerning economy and healthcare exhibit the highest coverage of scholarly references, while those related to law show the lowest.

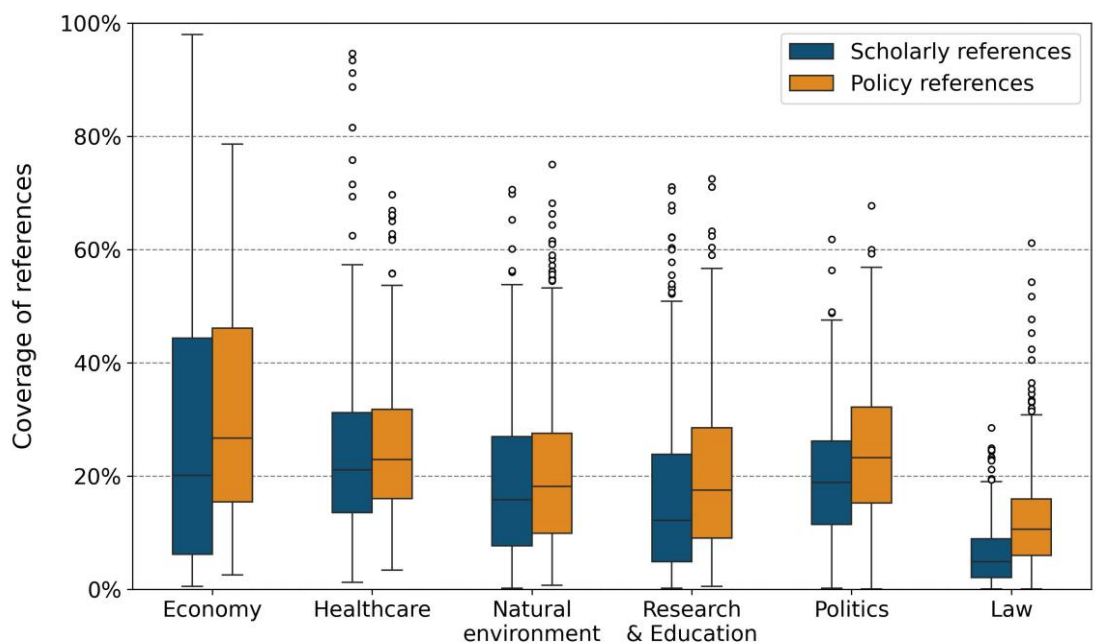

**Figure 9**. Distribution of the coverage of scholarly and policy references across topics within the six clusters.

When we consider policy references, Figure 8(c) shows that topics in the left part of the network, particularly those related to politics and law, become relatively more prominent. This suggests that policy documents in these areas are more likely to reference other policy documents rather than scholarly publications. These observations are consistent with the findings in Figure 9. Similar trends for the average number of scholarly and policy references across topics are displayed in Figures E1 and E2 in Appendix E.

## 4. Discussion

Amid the growing demand for science to benefit society, the scope of research evaluation has expanded beyond assessing solely the academic impact of research outputs to encompass their broader societal impact (Bornmann, 2013, 2014). Governments and funders expect researchers to justify the impact of their work by demonstrating "economic benefits, policy uptake, improved health and community outcomes, industry application and/or positive environmental effects" (Alla et al., 2017). As a specific form of societal impact, policy impact illustrates how research outputs provide concrete evidence to support policymaking processes, which can be reflected through references to research outputs within policy documents. From the policy perspective, referencing scientific evidence is a significant manifestation of science-based policymaking, increasingly considered important for "enhancing the quality and legitimacy of democratic governance systems" (Pedersen, 2014).

Against this background, Overton provides new opportunities to explore how policymaking seeks and utilizes scientific evidence and, conversely, how research outputs impact policy. This study presents an overview of the presence of both scholarly and policy references in Overton-indexed policy documents across various bibliographic dimensions, including publication years, policy sources, geographic regions, languages, subject areas, and topics. The results illuminate the potential uses and limitations of Overton for studying evidence-based policymaking and for offering evidence of the policy uptake of research outputs.

*4.1. Overton as a global comprehensive data source on policymaking*

Overton adopts a broad definition of policy documents, encompassing various policy sources such as governments, think tanks, IGOs, open repositories, and NGOs. This diversity allows for a multifaceted understanding of policymaking processes by incorporating perspectives from official governmental policies, independent research organizations, and international bodies. Researchers can choose to focus exclusively on policy documents issued by governments or specific policy sources within the Overton database, enabling more refined analyses of policymaking types and impacts.

However, an evident limitation, acknowledged by Overton itself, is that it cannot be expected to have indexed *all* policy documents ever written (Overton, 2024j). For example, there is a pronounced overrepresentation of policy documents published after 2009 and from developed countries such as the US and European nations. In contrast, regions like Asia, Africa, and Latin America are underrepresented, implying that the reference data extracted from Overton may predominantly reflect policy perspectives from certain regions. Overton (2021) explains several reasons for this geographic bias, including the concentration of think tanks and NGOs in specific countries, the varying degrees to which governments make their documents available online, and the different governmental levels that Overton tracks across countries. Consequently, observations based on the entire Overton database might exhibit a bias toward the policy focuses of particular geographic regions. Therefore, when utilizing the Overton database, it is important to recognize the different roles countries may play in shaping the policy landscape indexed in the database. On the other hand, the geographic information of policy sources provided by Overton also offers potential for research on differing policy focuses and the evolution of policies across countries.

The language distribution further accentuates this imbalance. Similar to many other bibliometric or altmetric data sources that exhibit a bias toward English-language sources (Hicks et al., 2015; Robinson-García et al., 2014), Overton currently includes policy documents in 74 languages, but English-language documents comprise 67.1% of the dataset and contribute 83.3% to the overall scholarly reference data and 82.1% to the policy reference data. This linguistic skewness implies a lack of diversity in the policy perspectives available for analysis in Overton, and may overlook culturally specific policymaking practices. Nevertheless, despite the predominance of English-language documents, those written in other languages are still present, offering opportunities to delve into public policy within local contexts. Future research should focus on addressing these linguistic and geographical biases, potentially identifying additional policy sources, and thus contributing to a more comprehensive and unbiased view of policymaking – a need also recognized by Overton itself (Overton, 2021). Meanwhile, researchers using Overton should be aware of this issue.

The subject area and topic information assigned by Overton aid in understanding the main themes of its indexed policy documents. These documents cover a wide range of subject areas and topics, with particular concentrations in science and technology, economy, politics, health, law, and environment. Future studies should investigate whether other important areas are missed by Overton – perhaps due to not indexing relevant policy documents related to these areas – or whether some areas have a more indirect relationship with policymaking.

*4.2. Overton as a source of science-policy and policy-policy interactions*

In this study, we focused closely on the presence of scholarly and policy references in Overton-indexed policy documents, as the inclusion of these reference data offers valuable opportunities to study the interactions between science and policy (i.e., science-policy interactions) and among policy documents themselves (policy-policy interactions). Overall, Overton provides substantial number of reference data, with the scholarly references being more extensive compared to Altmetric, as previously reported by Szomszor and Adie (2022). This allows researchers to understand how evidence informs policy decisions on a larger scale. Additionally, the availability of policy reference data in Overton enables investigations into the relationships among polices, and how they build on and influence each other.

Despite the substantial reference data available in Overton, our results indicate that the presence of scholarly and policy references varies significantly across different dimensions. Over time, there is a notable tendency of policy documents to cite other earlier policy documents. This is partly due to the increasing coverage of Overton, which facilitates the identification of references to previously indexed policy documents. Newly published policy documents have a longer time window and thus a larger data pool from which to search for previous policy documents as potential references. However, it remains an open question whether older policy documents also substantially cited other policy documents not indexed by Overton. Therefore, policy-to-policy citations are largely limited to more recent observations, coinciding with the advent of the Overton database. Consequently, strong conclusions on the "impact of policy on policy" cannot be extrapolated for earlier periods.

An opposite pattern can be observed for policy-to-science citations. Our results show that, over time, policy documents exhibit a much lower presence of citations to scientific research. This observation could result from the expansion of policy documents indexed by Overton, with an increasing number of documents that do not cite science, particularly those from governmental sources (Szomszor & Adie, 2022). Future research should delve into the reasons of this larger set of policy documents does not cite science, examining whether this is due to indexing policies, technical issues in citation matching, or a genuine decrease in science-based policymaking.

We found that policy documents from governments contribute the most to the overall reference data in Overton, largely due to the high volume of documents from this source type, followed by those from IGOs. This observation is consistent with the findings of Cristofoletti *et al.* (2023), who noted that governments and IGOs were the main contributors to policy citations of research funded by the São Paulo Research Foundation (FAPESP). However, in terms of the coverage and average of reference data, governmental documents cite fewer science or policy sources compared to other policy sources (i.e., think tanks, IGOs, and open repositories or NGOs). This suggests that these other organizations may place a greater emphasis on incorporating academic research into their policy documents, while government-sourced policy documents are less likely to cite academic research.

Two potential factors may explain the overall lower presence of reference data in government-sourced policy documents. First, government policymakers may exhibit a relatively low level of commitment to evidence-based decision-making. This could manifest either as a lack of emphasis on referencing and utilizing scientific evidence during the policymaking process, or as instances where scientific evidence is consulted but not explicitly cited in the final policy text. Future research could adopt qualitative methods, such as interviews, to explore policymakers' tendencies to reference scientific literature during decision-making or to investigate the motivations and processes underlying the practice of evidence-based policymaking. This approach would provide a more comprehensive understanding of the realities of evidence-based policymaking beyond bibliometric data, and help identify strategies to better integrate scientific evidence into policy formulation and implementation processes. Second, technical challenges may also contribute to the lower presence of reference data in government-sourced policy documents. In countries like Japan and China, where the identification of references in the Overton database is relatively low, most policy documents originate from local governments. These governments may predominantly use local languages and cite local literature, which complicates Overton's ability to accurately detect and identify the scientific literature referenced in these documents. This issue will be further elaborated in the following sections.

In terms of languages and geographic regions, since the methodology applied by Overton for detecting references works best on Western style references (Overton, 2024b), it is not surprising that policy documents written in Western languages (e.g., English or French) exhibit a higher presence of both scholarly and policy references. Moreover, because Overton's approach identifies scholarly references by matching reference strings with Crossref, it is likely more difficult for Overton to match references written in non-Western languages (Overton, 2024c) and references to scientific literature that lack DOIs, which are more commonly found in local scientific literature (Alonso-Alvarez & Van Eck, 2024; Maricato et al., 2023). As a result, Overton is more likely to miss references in non-Western policy documents if they are citing local sources or using special reference styles (Overton, 2024b), such as those in Chinese and Japanese. This language bias intensifies differences among countries as well. Policy documents from countries located in Europe, Oceania, and North America are more likely to contain scholarly or policy references, as are those originating from IGOs. By contrast, policy documents from Asian, African, and Latin American countries exhibit a much lower presence of references. These issues caution Overton users regarding the interpretation of policy data when different countries are considered. Future research should focus on minimizing these biases – for example, by identifying more local policy sources and improving the detection of references in different languages.

Based on the policy document citations provided by Altmetric, Haunschild and Bornmann (2017) found that the WoS-indexed papers in the fields of economics and certain areas of medicine and health sciences are comparatively more present among policy document citations than those from other fields. Noyons (2019) also reported that policy citations identified by Altmetric were primarily concentrated on WoS-indexed papers from the fields of social sciences, life and earth sciences, and biomedical sciences. In this study, our analysis of policy-to-science references supports the notion that policy topics related to economy and healthcare are strongly represented in the policy documents that cite more science. Other policy topics strongly based on science include natural environment and scientific research & education activities. These results support the idea that policy-science interactions are more prominent in these policy subject areas.

Policy-to-policy references, on the other hand, are also relatively common in the aforementioned policy areas that interact more with science. An important exception is documents on certain topics related to politics and law, which more frequently reference other policy documents. This is understandable, as these policy topics inherently have a strong policy component. Overall, variations in referencing patterns across policy subject areas suggest potentially different levels of interaction in the policymaking processes. In other word, there are varying degrees of science-based policymaking across policy subject areas, but we can also observe topics with a strong "policy-based" referencing component. In the practice of policy impact measurement based on Overton data, research outputs from the fields of economics, health sciences, and natural environment are more likely to be observed with policy uptake.

*4.3. Limitations of the study*

This study has several limitations. First, the version of the Overton database we analyzed is from June 2025. Given that Overton continuously updates its database with new policy documents from tracked sources and adds new policy sources (Overton, 2024f), the distribution of its indexed policy documents might change over time. Second, the policy sources defined by Overton refer to the locations where the policy documents were collected, but these are not always the same as the authors of the policy documents. Correspondingly, the geographic regions of policy documents indicate the location of the policy sources but not necessarily the geographic information of the authors or originators of the documents. This might cause deviations in the geographic distribution of policy documents, particularly for those retrieved from open repositories. Lastly, Overton is not free from data quality problems, which has been emphasized by Overton to promote responsible metrics (Overton, 2024j). For example, during the detection and matching of scholarly references, Overton prioritizes precision over recall, so it is possible to miss some references in policy documents (Overton, 2024c).

**5. Conclusions**

This study presents a large-scale analysis of the distribution of policy documents and the presence of reference data in the Overton database, providing a global and systematic understanding of the infrastructure of this new data source. Overton indexes a substantial number of policy documents, organized into categories and including multiple metadata elements, making it possible for future research to focus on more targeted research questions – for example, policy trends in climate change in the US, policy connections in the implementation of Sustainable Development Goals, or the use of scientific evidence in the development of COVID-19-related policies.

Moreover, approximately 7.7% of the indexed policy documents in Overton contain scholarly references, and 10.6% include policy references, underscoring the potential for further advancing evidence-based policymaking on a global scale. The reference data enables researchers to capture relevant interactions between policy and science and among policies themselves, which is valuable for improving the understanding of how scientific knowledge informs policymaking and, conversely, how policymakers access scientific information during their decision-making. However, it should be noted that there are coverage issues in Overton regarding certain time periods, geographic regions, languages, and topics. The overrepresentation of certain policy sources, countries, and topics suggests that conclusions drawn from the Overton database may disproportionately reflect the policymaking practices of specific types of policymakers, world regions, and subject areas. This limitation necessitates caution when generalizing results and underscores the need for efforts to expand the database's coverage to include more comprehensive policy documents from underrepresented entities.

**Funding information**

This work was supported by the National Natural Science Foundation of China (72304274 to Z.F.), the China Scholarship Council (202206840041 to B.M.), and the South African DSI-NRF Centre of Excellence in Scientometrics and Science, Technology and Innovation Policy (SciSTIP) (to R.C.).

**Acknowledgments**

The authors thank Overton for providing the data for research purposes, and thank the anonymous reviewers for their valuable suggestions.

**Declaration of interests**

The authors declare that they have no conflicts of interest.

**Data availability**

The data underlying this study are subject to access restrictions based on the contract with Overton and therefore cannot be shared publicly. Researchers interested in accessing the dataset can do so through the Overton platform, in accordance with their subscription and access policies. For further details, please refer to Overton's website: https://www.overton.io/.

**Appendix A**

A total of 6,918,404 unique cited DOIs were detected by Overton across its indexed 17.5 million policy documents. Table A1 presents the coverage of these cited DOIs in four academic databases: the Web of Science (70.2%), Scopus (64.8%), Dimensions (92.0%), and OpenAlex (95.2%). It is important to note that the CWTS in-house Scopus database primarily contains records from post-1995, which explains the slightly lower coverage in Scopus compared to the Web of Science, as DOIs from before 1996 were unavailable for matching. After combining the results, 6,591,099 unique DOIs were indexed in at least one of the four academic databases, representing 95.3% of all cited DOIs identified by Overton.

**Table A1**. Coverage of cited DOIs in policy documents across four academic databases

| Academic database | Version | Number of DOIs indexed | Proportion of DOIs indexed |
|---|---|---|---|
| Web of Science | March 2025 | 4,855,779 | 70.2% |
| Scopus | March 2025 | 4,482,697 | 64.8% |
| Dimensions | July 2024 | 6,367,057 | 92.0% |
| OpenAlex | August 2024 | 6,584,032 | 95.2% |

**Appendix B**

Figure B1 illustrates the proportional composition of policy documents published by the four source types over time. Overton indexes an increasing share of government policy documents, which have become the predominant source in the database. In contrast, the proportion of IGO documents has gradually declined, reaching a level comparable to that of think tanks in recent years. Other sources contribute only a small fraction of the policy documents indexed in Overton.

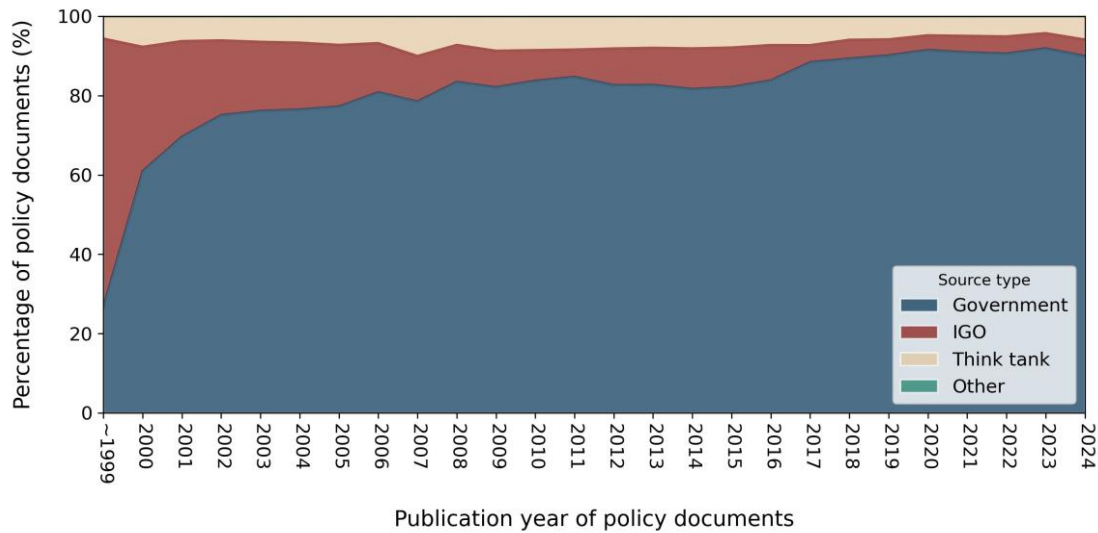

**Figure B1**. Proportion of Overton-indexed policy documents by source type across publication years.

**Appendix C**

Figure C1 presents the distribution of Overton-indexed policy documents by country, grouped by six continents. North American countries have the highest number of indexed policy documents in Overton, followed by IGOs and European countries. In contrast, countries in Latin America & Caribbean, Asia, Oceania, and Africa generally have fewer policy documents indexed. Notable regional disparities exist, with specific countries such as Japan, Australia, and South Africa standing out as exceptions due to their relatively high number of Overton-indexed policy documents.

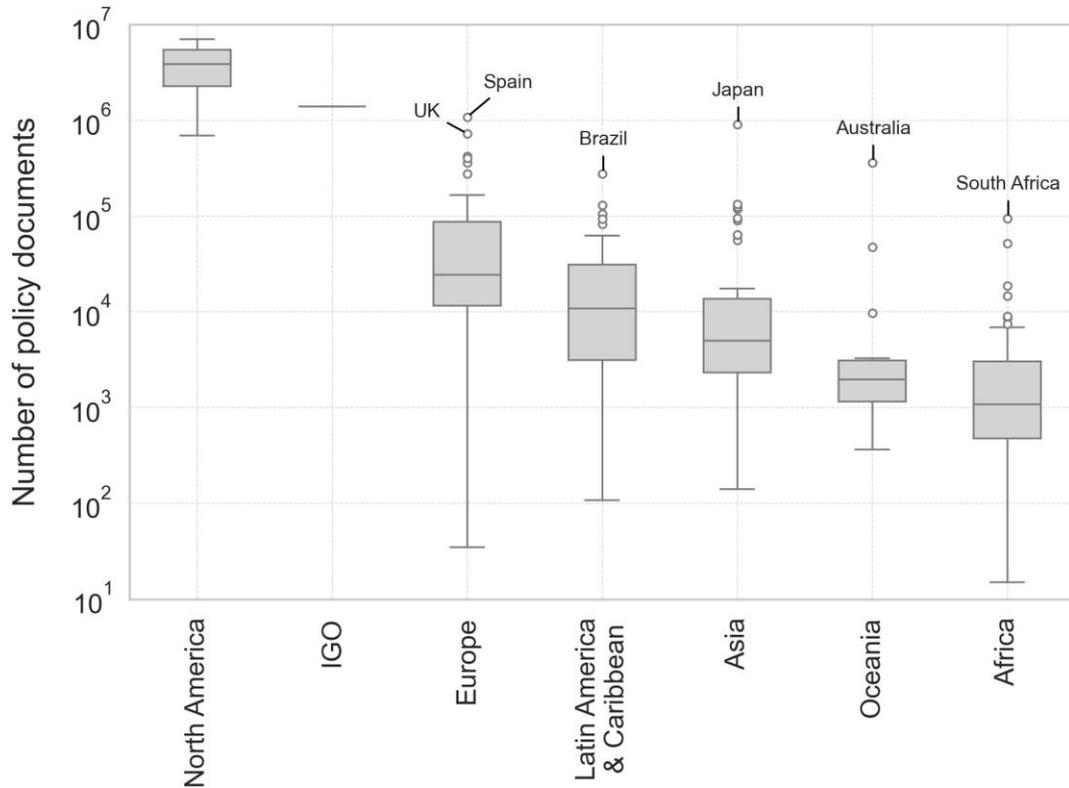

**Figure C1**. Distribution of Overton-indexed policy documents by country across six continents, with IGOs categorized separately.

**Appendix D**

Figure D1 presents, for each of the 18 subject areas, the proportions of policy documents associated with the six topic clusters identified in Figure 8(a). Since Overton allows a single policy document to be assigned to multiple topics, each subject area includes documents linked to all six clusters, albeit to varying degrees. Most subject areas are more strongly associated with their thematically related topics. For example, topics concerning the "natural environment" account for the largest share of policy documents in the areas of *Environment*, *Weather*, *Disaster, accident and emergency incident*. Similarly, topics related to "research & education" dominate the areas of *Science and technology*, *Education*, and *Arts, culture and entertainment*. This pattern highlights both the alignment between subject areas and document topics, as well as the diversity of topics represented within each subject area.

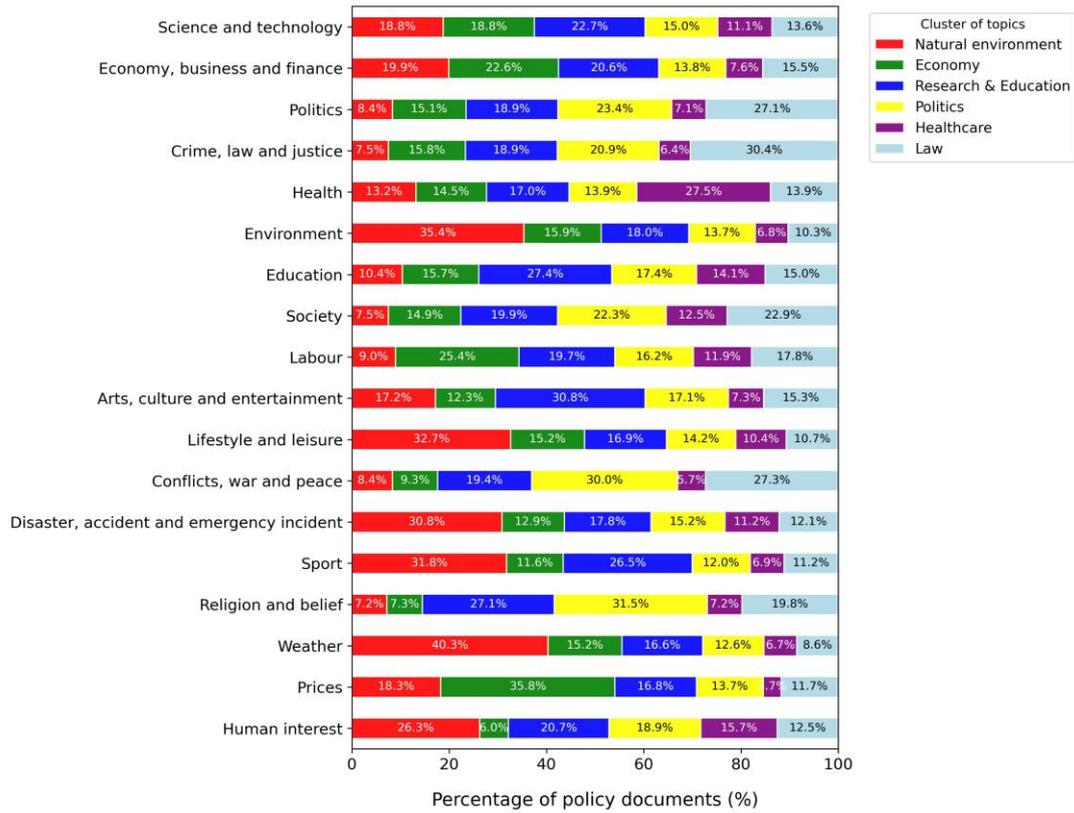

**Figure D1**. Proportion of policy documents related to six topic clusters across 18 subject areas.

**Appendix E**

Figure E1 presents an overlay visualization of the co-word network of topics, scored by the average number of scholarly or policy references in the related policy documents. The overall patterns are similar to those observed in Figure 8, which is based on coverage values. Policy documents related to healthcare, natural environment, economy, and research stand out in terms of scholarly references. However, when it comes to policy references, topics concerning politics and law tend to rely more heavily on citing other policy documents. These findings are consistent with the trends illustrated in Figure E2, which shows a boxplot of the distribution of the average of references across topics within the six distinct clusters.

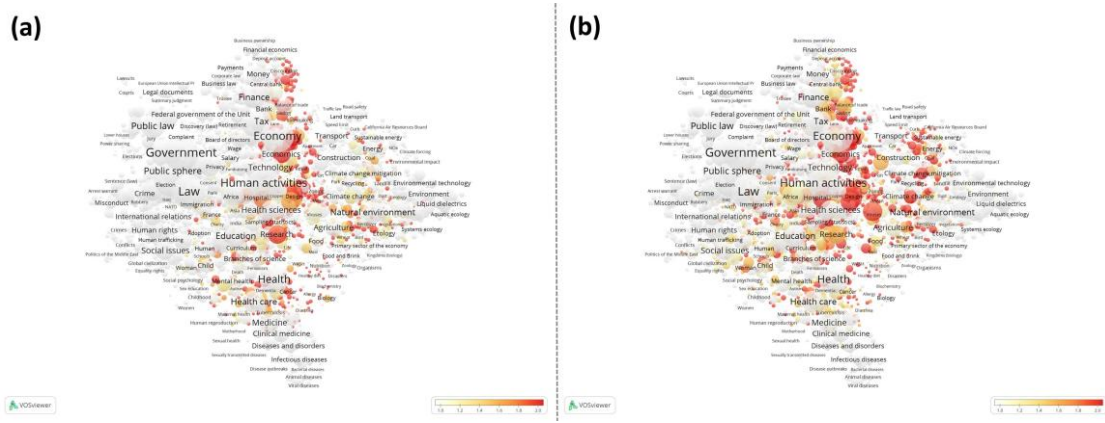

**Figure E1.** Overlay visualizations of the co-word network of topics, colored based on the values of (a) average of scholarly references in related policy documents and (b) average of policy references in related policy documents. The scores of topics in the overlay visualizations are normalized using the "divide by mean" function in VOSviewer.

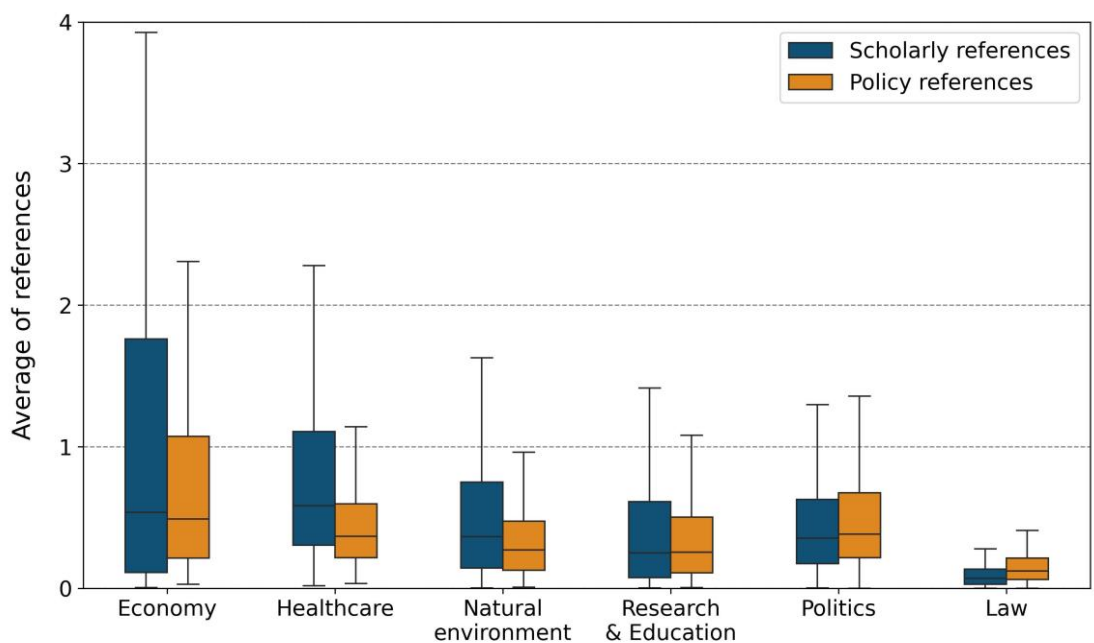

**Figure E2.** Distribution of the average of scholarly and policy references across topics within the six clusters. For clarity, outliers are removed from the visualization to provide a more concise representation of the general trends.